\documentclass[aps,prl,nobalancelastpage,superscriptaddress,twocolumn,longbibliography,nobibnotes]{revtex4-1}

\usepackage[utf8]{inputenc}
\usepackage[american,british]{babel}
\usepackage[T1]{fontenc}
\usepackage[pdftex]{graphicx}  
\usepackage{xcolor}
\usepackage{dcolumn}
\usepackage{physics}
\usepackage{braket}
\usepackage{bm}
\usepackage{amsmath,amsthm,amssymb}
\usepackage{color}
\usepackage{verbatim}
\usepackage[normalem]{ulem}

\newcommand{\lm}[1]{\textcolor{teal}{#1}} 

\newcommand{\Ll}{\mathcal{L}}

\newcommand{\lptms}{Universit\'e Paris-Saclay, CNRS, LPTMS, 91405, Orsay, France}
\newcommand{\iuf}{Institut Universitaire de France, 75005 Paris, France}
\newcommand{\ningbo}{Institute of Fundamental Physics and Quantum Technology, and
School of Physical Science and Technology, Ningbo University, Ningbo, Zhejiang 315211, China}
 
\newcommand{\smt}{Science, Mathematics and Technology Cluster, Singapore University of Technology and Design, 8 Somapah Road, 487372 Singapore}
\newcommand{\epd}{Engineering Product Development Pillar, Singapore University of Technology and Design, 8 Somapah Road, 487372 Singapore}
\newcommand{\cqt}{Centre for Quantum Technologies, National University of Singapore 117543, Singapore}
\newcommand{\inns}{Institut f\"ur Quantenoptik und Quanteninformation, \"Osterreichische Akademie der Wissenschaften, Technikerstraße 21a, 6020 Innsbruck, Austria} 
\newcommand{\uinn}{Institut f\"ur Experimentalphysik, Universit\"at Innsbruck, Technikerstraße 25, 6020 Innsbruck, Austria}

\usepackage{hyperref}
\hypersetup{
 colorlinks=true,
 linkcolor=blue,
 anchorcolor = blue,
 citecolor = blue,
 filecolor = blue,
 urlcolor = blue
}

\def \be {\begin{equation}} 
\def \ee {\end{equation}}

\def \la {\langle} 
\def \ra {\rangle}

\begin{document}

\author{Jiaozi Wang} 
\thanks{These authors contributed equally.} 
\affiliation{\ningbo} 

\author{Manoj K. Joshi} 
\thanks{These authors contributed equally.} 
\affiliation{\smt} 
\affiliation{\inns} 

\author{Luca Capizzi} 
\affiliation{\lptms}

\author{Rainer Blatt} 
\affiliation{\inns} 
\affiliation{\uinn}  

\author{Christian F. Roos} 
\affiliation{\inns} 
\affiliation{\uinn}   

\author{Leonardo Mazza}
\email{leonardo.mazza@universite-paris-saclay.fr}
\affiliation{\lptms}
\affiliation{\iuf}

\author{Dario Poletti} 
\email{dario\_poletti@sutd.edu.sg }
\affiliation{\smt}
\affiliation{\epd}
\affiliation{\cqt}

\title{Observation of universal hierarchical relaxation in a quantum simulator}

\date{\today}

\begin{abstract}
Autocorrelation functions play a key role in the theoretical characterization of the dynamical properties of interacting many-body quantum systems. 
Recently, bringing together the eigenstate thermalization hypothesis and hydrodynamics, it was theoretically predicted that the relaxation of autocorrelators can be described by the \textit{relaxation-overlap inequality}, which, when saturated, predicts a hierarchy of relaxation exponents for a set of operators that are easily identified and constructed.  
Here, we employ a trapped-ion quantum simulator to experimentally demonstrate it in the diffusive regime; 
to do so, we have extended the theory of the overlap-relaxation inequality to systems with multiple conservation laws.
Our study thus opens the path to a thorough characterization of the relaxation to equilibrium and appearance of hydrodynamic behavior in quantum matter.   
\end{abstract}

\maketitle 

\paragraph{\textbf{Introduction ---}}

Understanding the equilibration of quantum systems is one of the most important problems in many-body physics
and several theoretical frameworks have been proposed to describe the late-time dynamics of quantum systems~\cite{Silva_review_RMP, Mori_2018}.
The \textit{eigenstate thermalization hypothesis} (ETH), for instance, describes thermalization in isolated quantum systems by treating the Hamiltonian eigenstates at a given energy density as random vectors~\cite{berry1977, Deutsch-91, Srednicki-99};
this simple assumption is enough to derive the emergence of thermal behavior from the unitary quantum dynamics of generic chaotic systems~\cite{rdo-08,Rigol-09,sr-10,rs-10,bkl-10,krrg-12,ghl-12,bmh-14,bmh-15,kpsr-13,shp-13,kih-14,skngg-14,ksg-15,mfsr-16}.
Another quantitative framework that is routinely employed is \textit{hydrodynamics};
by proposing a coarse-grained view based on fluid cells that are locally at equilibrium, it describes the local spreading of the density of observables that are globally conserved, such as energy  and particle number~\cite{km-63,bkv-05,baw-06,msf-09,Spohn-12,lmmr-14,ngtsm-15,ttgp-15,mknsm-16,Crossno-16,lf-18,cdy-16,bcdf-16,dbd-19,Doyon-22}.

Bridging and integrating these two approaches, ETH and hydrodynamics, has produced interesting insights into quantum many-body physics~\cite{BalachandranPoletti2021, BalachandranPoletti2023, BalachandranPoletti2023b, cwxmp-24, WangPappalardi2026}, and classical too \cite{WangMazza2026}. 
For instance, it has been shown that the algebraic decay in time as $\sim t^{-\nu}$ of the autocorrelation functions at finite temperature of a local observable $\hat{\mathcal O}$, where $\nu$ is the relaxation exponent, 
satisfies the \textit{relaxation-overlap inequality} (ROI)~\cite{cwxmp-24}
\begin{equation}
\label{Eq:ROI}
 \nu \leq \frac{\mu d}{z}.
\end{equation}
Here, $z$ is the dynamical critical exponent, crucial in  any hydrodynamic theory, $d$ is the dimensionality of the system, and $\mu$ is the \textit{overlap order}, a quantity 
that is properly explained later in this letter, and is related to how the operator in the autocorrelator overlaps with local conserved quantities.
The fact that the large majority of the numerical and experimental studies has so far focused on observables with $\mu=1$ has hidden this intriguing dependence. 
The relaxation is universal in the sense that it  depends only on the overlap order,  rather than on the details of the considered operator. Furthermore, it is also hierarchical, that is, given an operator with an algebraic relaxation, and a setup that saturates this inequality, one can construct a hierarchy of operators with faster and faster relaxation. 

The exceptional blooming of quantum simulators allowing for the controlled investigation of synthetic quantum matter with unprecedented precision, even to study the emergence of hydrodynamic behavior \cite{ZuYao2021, Joshi-22, PengCappellaro2023, WeiZeher2022, RosenbergRoushan2024, ShiFan2024, WienandBloch2024, ZhangPoletti2024}, makes testing the ROI in an experimental device a very intriguing perspective. 
However, the ROI currently rests on the assumption that energy is the only conserved quantity of the model and thus it cannot be readily employed to interpret the data of most existing platforms for quantum simulation.
Indeed, the models that are routinely used to describe those devices often feature another conserved quantity, typically interpreted as the number of particles or the magnetization along one direction. 

\begin{figure}
	\includegraphics[width=\columnwidth]{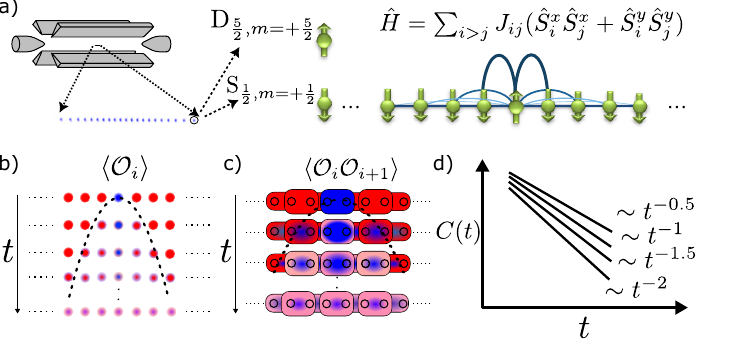}
 \caption{$(a)$ Sketch of the ion trap used for the experiment, which zooms into an image of the ions, and a depiction of the long-range interactions. (b) Propagation of a single-site and c) two-site observable. (d) Depiction of hierarchy of decays of the autocorrelation functions. }
\label{fig:summary}
\end{figure}

In this Letter, we experimentally investigate the emergence of this universal and hierarchical relaxation of autocorrelation functions in a trapped-ion quantum simulator, reanalyzing data produced by the experiment reported in ~\cite{Joshi-22}, while developing an appropriate generalization of the ROI for systems with several conservation laws. 
Our work thus opens the path to a thorough and quantitative characterization of autocorrelation functions and of relaxation dynamics in quantum many-body systems. 
A sketch of some key aspects of this work is presented in Fig.~\ref{fig:summary}. Fig.~\ref{fig:summary}(a) shows the trapped ions setup, which has interactions beyond the nearest neighbour; in panels (b,c), we pictorially depict the spreading of single-site and two-site operators respectively, which is constrained by the transport of local conserved quantities. In Fig.~\ref{fig:summary}(d) we illustrate that correlators for different operators can decay algebraically with different exponents, a signature of the hierarchical nature of thermalization.          

\paragraph{\textbf{Experimental setup and Hamiltonian ---}}

We set up a one-dimensional chain of $^{40}$Ca$^+$ ions, with two electronic states $\ket{S_{1/2},m=+1/2}$ as spin-down $\ket{\downarrow}$ and $\ket{D_{5/2}, m=+5/2}$ as spin-up $\ket{\uparrow}$, see Fig.~\ref{fig:summary}(a). 
We initialize the ion chain in randomly chosen product states of spin-up and spin-down states using individual-ion addressing. Averaging over many such product-state realizations realizes an effective infinite-temperature ensemble for the present study. 
The spin-spin interactions are generated by a two-tone laser beam that approximately induces an Ising-type interaction with an approximate power-law decay. The two-tone laser off\lm{-}resonantly couples all the transverse radial motion and electronic degrees of freedom of the ion chain. 
We shift the center of the two tones by a large amount such that it energetically penalizes excitations that do not conserve the total magnetization, and the interaction in rotating frame is expressed as long range XX interaction. 
With this, we produce a model whose effective dynamics is described by the long-range XX Hamiltonian, also depicted in Fig.~\ref{fig:summary}(a), 
\begin{equation}\label{eq:Ising_H}
\hat H=\sum_{i<j}J_{ij}
\left(\hat S^{x}_{i} \hat S^{x}_{j}+
\hat S^{y}_{i} \hat S^{y}_{j} \right) 
\quad \; J_{ij} = \frac{J_0}{|i-j|^{\alpha}}.
\end{equation}
Here, $\hat S^a_i$ ($a=x,y,z$) are 
Pauli 
operators at lattice site $i$, $L$ is the number of spins. $J_0$ and $\alpha$ are the coupling strength and range of interaction, respectively, which are tunable in the system. 
We note that there is also a transverse field term arising as an artefact of the rotating-wave approximation and the motional spectral structure of trapped ions. 
In trapped-ion experiments, this term can be effectively transformed into site-dependent $\hat S^{z}_{i}$ terms using additional light-shift operations \cite{Joshi-22}, making its effect negligible for the present study.
Besides the energy, the model in Eq.~\eqref{eq:Ising_H} conserves the total magnetization along the $z$-axis, $\hat S^z = \sum_j \hat S^z_j$. 
This setup was recently used to characterize experimentally the connected correlation functions 
$\la \hat S^z_j(t) \hat S^z_j\ra_c$ 
of the infinite-temperature state and validate the predictions from hydrodynamics theory for a classical long-ranged diffusion process. 
Specifically, it has been shown that at late times, before approaching thermal equilibrium, the autocorrelation function of $\hat S^z_j$ displays a hydrodynamic tail $\sim t^{- \nu}$, and that $\nu$ is determined by $\alpha$, the parameter that sets the range of the coupling in Eq.~\eqref{eq:Ising_H}. In particular, for $\alpha \geq 3/2$ the exponent is $\nu=1/2$, corresponding to normal diffusion, while for $1<\alpha<3/2$ the system is superdiffusive with $\nu=1/(2\alpha-1)$.      
In this work we will focus on the case $\alpha=3/2$; 
importantly, we will only consider observables that saturate the ROI.

\paragraph{\textbf{ETH and ROI in presence of more than one conserved quantity ---}} 
The difficulty of applying ETH to Hamiltonian~\eqref{eq:Ising_H} is that the model conserves both energy and the $z$ component of the magnetization. 
Extensions of ETH in the presence of global symmetries (such as $U(1)$, relevant in this case) have been proposed in Refs.~\cite{Belin-22,mbisy-23, KranzlYungerHalpern2023, LasekYungerHalpern2026, MajidiYungerHalpern2026, PatilRigol_2025}, and amount to postulating ETH in each symmetry sector.
Here, we propose a similar ansatz, assuming that a \textit{finite}  family of $N$ \textit{extensive} conserved charges $\{\hat Q_a \}_{a=1,\dots,N}$ is present.
We stress that the number $N$ should not increase in the thermodynamic limit, as it happens in integrable systems, which are thus not described by the present theory \footnote{We note, however, that the relaxation-overlap inequality is expected to be valid also in integrable many-body systems which have a generic spectrum \cite{BalachandranPoletti2023}.}. 
Additionally, we highlight that restricting to extensive charges means that a writing of the form $\hat Q_a = \sum_j \hat q_{a,j}$ is possible, with $\hat q_{a,j}$ 
a local operator supported around the lattice site $j$, or extending over the entire chain but with an appropriately fast decay. 
For notational convenience, we represent these charges with an $N$-component operator-valued vector $\hat{\mathbf{Q}}$, and, because of its relevance to our setup, we consider that all charges pairwise commute, $[\hat Q_a, \hat Q_{a'}]=0$.
In fact, for the Hamiltonian Eq.~\eqref{eq:Ising_H}, there are two independent charges, $N=2$, with the first charge being the Hamiltonian, $\hat Q_1 = \hat H$, and the second being the magnetization along $z$, $\hat Q_2 = \hat S^z$, and the two charges commute.

We consider the Hilbert space of the quantum states of the spin chain and a basis composed of common eigenvectors of the conserved charges 
$\{Q_a\}_{a=1}^N$, noted $\{\ket{\mathbf{Q}_j}\}$ and labeled by the integer $j$; this basis satisfies the defining relation $\hat {\bf Q} \ket{\mathbf Q_j} = \mathbf Q_j \ket{\mathbf Q_j}$.
Note that no degeneracy in $\mathbf Q_j$ is assumed, hence for $i \neq j$ the vectors $\mathbf Q_i$ and $\mathbf{Q}_j$ are different.
Given a local observable $\hat {\mathcal{O}}$, we postulate, in the limit of large systems, the generalized ETH relation 
\begin{equation}\label{eq:ETH_ansatz}
\bra{\mathbf Q_i}\hat{\mathcal{O}}\ket{\mathbf Q_j} = \delta_{ij} \mathcal{O}(\mathbf{q}) + \frac{1}{\sqrt{\rho(\mathbf q)}}f_{\mathcal{O}}({\mathbf{q}},\boldsymbol{\omega}) R_{ij},
\end{equation}
which is expressed in terms of the generalized charge-density vector $\mathbf q$ and
the generalized energy-difference vector $\boldsymbol{\omega} $, defined as
\begin{equation}
 \mathbf{q} :=(\mathbf{Q}_i+\mathbf{Q}_j)/2L ; \quad
 \boldsymbol{\omega} :=\mathbf{Q}_i- \mathbf{Q}_j.
\end{equation}
Moreover, $\rho(\mathbf{q})$ is the density of states at a given charge density $\mathbf{q}$, and $R_{ij}$ is a random matrix with zero mean and unit variance.

The diagonal part of the expression in~\eqref{eq:ETH_ansatz} corresponds to the expectation value of $\hat{\mathcal{O}}$ in a generalized Gibbs state with generalized vector temperature
$\boldsymbol{\beta}$, namely
$\rho_{\boldsymbol{\beta}} \sim e^{-\boldsymbol{\beta} \cdot \hat {\mathbf Q}}$.
In particular, one can use the expression in Eq.~\eqref{eq:ETH_ansatz} to show that in the thermodynamic limit the following expression holds
\begin{equation}
\label{eq:canonical}
\langle \hat{\mathcal{O}}\rangle_{\boldsymbol{\beta}}
:= \frac{\text{Tr}\left[ e^{-\boldsymbol{\beta} \cdot \hat{\mathbf{Q}}}\hat{\mathcal{O}}\right]}{\text{Tr}\left[ e^{-\boldsymbol{\beta} \cdot \hat{\mathbf{ Q}}}\right]} = \hat{\mathcal{O}}(\mathbf{q}(\boldsymbol{\beta})),
\end{equation}
stating the equivalence of the microcanonical expectation value~\eqref{eq:ETH_ansatz} with the canonical ensemble~\eqref{eq:canonical}.
The relation $\mathbf{q} \leftrightarrow \boldsymbol{\beta}$, relating
the microcanonical charge density to the canonical temperature, is given by $\boldsymbol{\beta} = dS(\mathbf q)/d\mathbf{q}$.

\begin{figure*}[t]
    \centering
    \includegraphics[width=\textwidth]{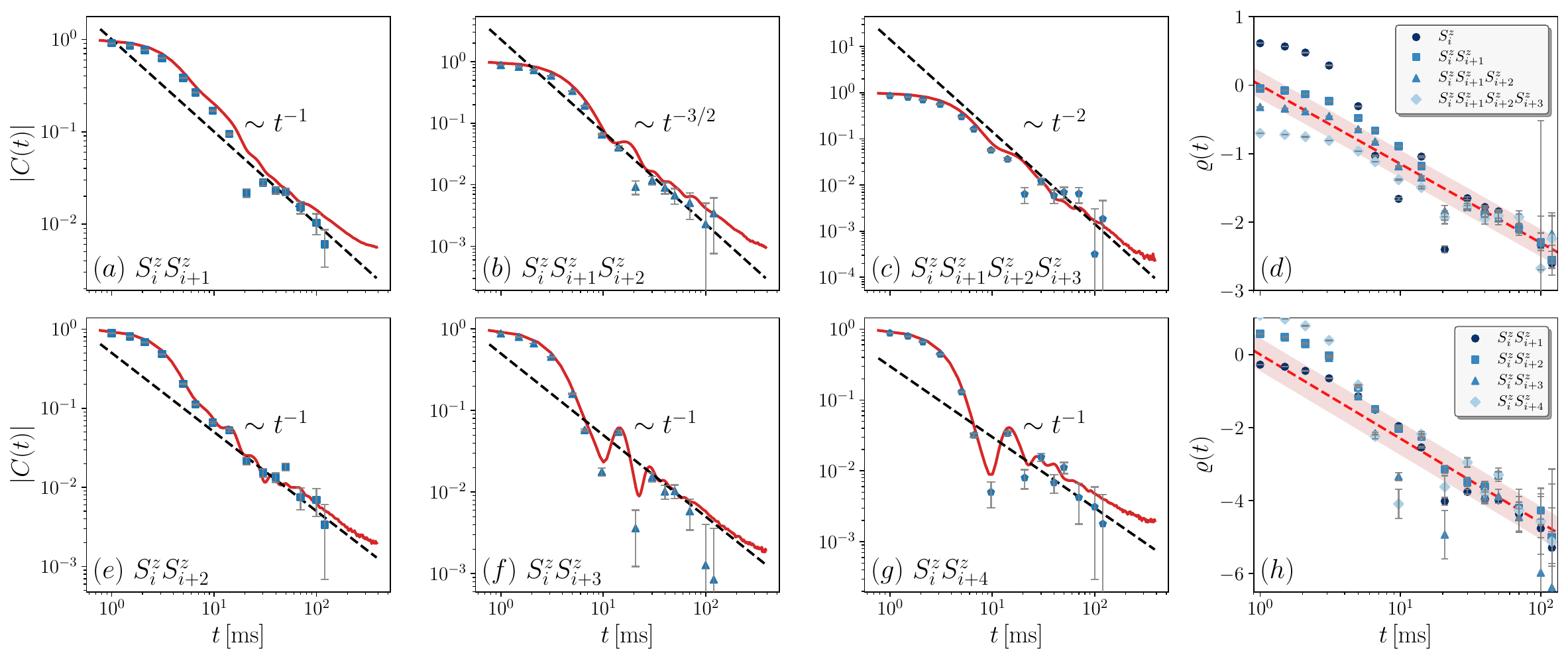}
    \caption{Auto-correlation function $C(t)$ for different observables in [(a-c)] and [(e-g)]. Blue dots, red solid lines and black dashed lines indicate the experimental data, numerical results and the analytical predictions, respectively. 
    The parameters of the Hamiltonian (\ref{eq:Ising_H}) used in the experiment and simulations are $J_0=232\; \rm{rad}/s$, $\alpha=1.5$ and $L=25$. 
    (d) Collapse of data for observables
$\hat{S}_{i}^{z},\hat{S}_{i}^{z}\hat{S}_{i+1}^{z},\hat{S}_{i}^{z}\hat{S}_{i+1}^{z}\hat{S}_{i+2}^{z},\hat{S}_{i}^{z}\hat{S}_{i+1}^{z}\hat{S}_{i+2}^{z}\hat{S}_{i+3}^{z}$, represented using $\varrho(t)$ from Eq.~(\ref{eq:ratio}); 
(h) Collapse of data for observables
$\hat{S}_{i}^{z}\hat{S}_{i+1}^{z},\hat{S}_{i}^{z}\hat{S}_{i+2}^{z},\hat{S}_{i}^{z}\hat{S}_{i+3}^{z},\hat{S}_{i}^{z}\hat{S}_{i+4}^{z}$, as a function of $\log{C(t)}$. 
For details about the numerical computations, see End Matters.  
    }
    \label{fig:combined}
\end{figure*}

Given a local observable $\hat{\mathcal O}$, on the basis of the previous results on relaxation in \cite{cwxmp-24}, we can assume that it displays an autocorrelator with algebraic relaxation to zero in the infinite volume limit, $\la \hat{\mathcal{O}}(t)\hat{\mathcal{O}}^\dagger\ra_{c} \sim t^{-\nu}$, with $\nu$ an exponent that depends on the operator $\hat{\mathcal{O}}$.
On the other hand, at finite system size $L$ and large time $t$, a plateau is theoretically expected to appear
\begin{equation}\label{eq:plateau}
\la \hat{\mathcal{O}}(t)
\hat{\mathcal{O}}^\dagger\ra_{c} \sim \frac{1}{L^\mu}. 
\end{equation}
When considering a $d$-dimensional system, the plateau value is $\left(L^d\right)^{-\mu}$. 

The exponent $\mu$ is dubbed  \textit{overlap order} and was anticipated in Eq.~\eqref{Eq:ROI}: it is defined as the maximal positive integer such that
\begin{equation}\label{eq:overlaps}
\la \hat{\mathcal{O}}\hat Q_{a_1}\dots \hat Q_{a_{m'}}\ra_{\boldsymbol{\beta},c} =0, \quad 1\leq m'\leq \mu-1,
\end{equation}
for any choice of the indices $a_j \in \{1,\dots, N\}$.
The name follows from the fact that the expectation values in Eq.~\eqref{eq:overlaps} can be interpreted as overlaps between the observable and the charges, and are directly related to the profiles of the expectation values at equilibrium. 
In particular, one can express them as $\la \hat {\mathcal{O}}\hat{Q}_{a_1} \dots \hat{Q}_{a_{m'}}\ra_{\boldsymbol{\beta},c} = (-1)^{m'}\partial_{\beta_{a_1}} \dots \partial_{\beta_{a_{m'}}} \la \mathcal{O}\ra_{\boldsymbol{\beta}} $; 
in other words, the partial derivatives of $\langle \mathcal{O} \rangle_{\boldsymbol{\beta}}$ vanish up to order $\mu-1$.

The relation between $\nu$ and $\mu$ follows from one further assumption, namely that the decay of the autocorrelator is monotonic (up to possible transient oscillations), as it is observed in numerical simulations in \cite{cwxmp-24}.
In a similar situation, when the system is finite and has length $L$, finite-size effects typically appear at time $J_0 t \sim L^{z_{\text{min}}}$,
where $z_{\text{min}}$ represents the dynamical critical exponent associated with the fastest hydrodynamic mode of the system. 
Indeed, in the presence of several conserved charges, different densities may spread according to different laws: here, to produce a bound, we select the fastest. 
For instance, $z=1$ is associated to a ballistic spreading whereas $z=2$ to a diffusive one; superdiffusive behaviors $1 < z < 2$ are possible in special situations (i.e.~long-range couplings). 
Since at $t \sim L^{z_{\rm min}}$ the autocorrelator decaying as $t^{- \nu}$ must be larger than or equal to the finite-size value $L^{- \mu}$, we can deduce the generalized ROI, which in $d$ dimensions is given by 
\begin{equation}\label{eq:rel_overlap}
\nu \leq \frac{d\;\mu}{z_{\text{min}}}; 
\end{equation}
this is the expression needed to account for the existence of multiple additional charges \cite{fn_overlaplongtime}. 

\paragraph{\textbf{Hierarchy of relaxation decay regimes ---}} 
While the relaxation dynamics of single-site operators has been explored experimentally with the quantum simulator described above in~\cite{Joshi-22}, this only gives a partial picture of how the system relaxes. Operators extending over more sites can still decay algebraically but with different (larger) exponents, following the ROI.   
In this work, we are going to experimentally probe this. We start by verifying the hierarchy of algebraic decays by considering the following operators $ \hat{\mathcal O}_{m,i} $
\begin{eqnarray}
 \hat{\mathcal O}_{1,i} = \hat S^z_i, &\qquad&
 \hat{\mathcal O}_{2,i} = \hat S^z_i \hat S^z_{i+1}, \nonumber \\
 \hat{\mathcal O}_{3,i} = \hat S^z_i \hat S^z_{i+1} \hat S^z_{i+2}, &\quad&
 \hat{\mathcal O}_{4,i} = \hat S^z_i \hat S^z_{i+1} \hat S^z_{i+2} \hat S^z_{i+3}.
 \label{Eq:4:Operators}
\end{eqnarray}
The operators $\hat{\mathcal O}_{m,i}$, for $m>1$,  are not the densities of an extensive conserved quantity (as it happens for $\hat{\mathcal O}_{1,i}$) and thus do not have their own hydrodynamic description.
In fact, at infinite temperature and zero magnetization, $O_{m,i}$ has overlap order $\mu=m$ with the conserved quantity $\hat{S}^z$, while $O_{2m,i}$ has overlap order $\mu=2m$ with $\hat{H}$. 
Hence, for odd values of $m$ the only conserved quantity that needs to be considered is $\hat{S}^z$, or a combination of even powers of $\hat{H}$ and odd powers of $\hat{S}^z$, while for even ones, the operator has the same overlap order with both $\hat{S}^z$ and $\hat{H}$. 
Local energy also diffuses \cite{NishikawaSaito2025}, thus not changing qualitatively the transport predicted by overlap with local magnetization. 
Our numerical results, although limited by finite-size effects, are overall consistent with the prediction of Ref.~[57] (see End Matter). Hence, we can qualitatively predict the relaxation dynamics for our choice of $\hat{\mathcal O}_{m,i}$ solely using the overlap order with $\hat{S}^z$ and its transport property.

We prepare the experimental system in different product states that are eigenstates of the $\hat S^z_j$ observables, with total magnetization as close as possible to zero, 
we let it evolve under the action of the Hamiltonian (\ref{eq:Ising_H}), and then we measure $\hat S^z_j$ on each site. This results in pairs of bit-strings (initial configuration and measured configuration), which can be used to evaluate  
\begin{equation} 
C_{m}(t)=\overline{\langle\mathcal{O}_{m,i}(t)\mathcal{O}_{m,i}(0)\rangle_{c}}  
\end{equation}
for the operators in Eq.~\eqref{Eq:4:Operators}, where the overline acting on a site-dependent quantity,  $\overline{\cdot}$, implies average over the sites of the setup.  
The experimental results are shown in Fig.~\ref{fig:combined}$(a$-$c)$.
The presence of an algebraic late-time tail is well highlighted by the plots in log-log scale, 
and it matches well the theoretical late-time behavior 
\begin{equation}
C_m(t)
\sim t^{-m/2}; \quad m = 2,3,4, 
\label{Eq:Power:Law}
\end{equation}
predicted by the saturated generalized ROI, Eq.~(\ref{eq:rel_overlap}), as our diffusive system has $z_{\rm min}=2$ and $d=1$. 
This curve is reported in the plots as a dashed line and this power-law has a high descriptive power, extending significantly the insights on relaxation presented in Ref.~\cite{Joshi-22} for the autocorrelation of $\hat{\mathcal O}_{1,i}$. 

To further confirm the hierarchy of algebraic relaxations, we consider a time $t^*$ such that the relaxation is in the algebraic regime for all the observables (i.e.~after the initial transient and before finite size effects), for which the function 
\begin{align}
    \varrho(t) = \frac 1 m \log\left(\frac{C_m(t)}{C_m(t^*)}\right) \label{eq:ratio}
\end{align}
should show a collapse for all the different observables, see Fig.~\ref{fig:combined}$(d)$.
While at early times the four curves (we also include $m=1$) are scattered, with the development of the hydrodynamic behaviour the collapse appears, highlighting the universal character of the hierarchy of relaxations.

\begin{figure}[tb]
	\includegraphics[width=\columnwidth]{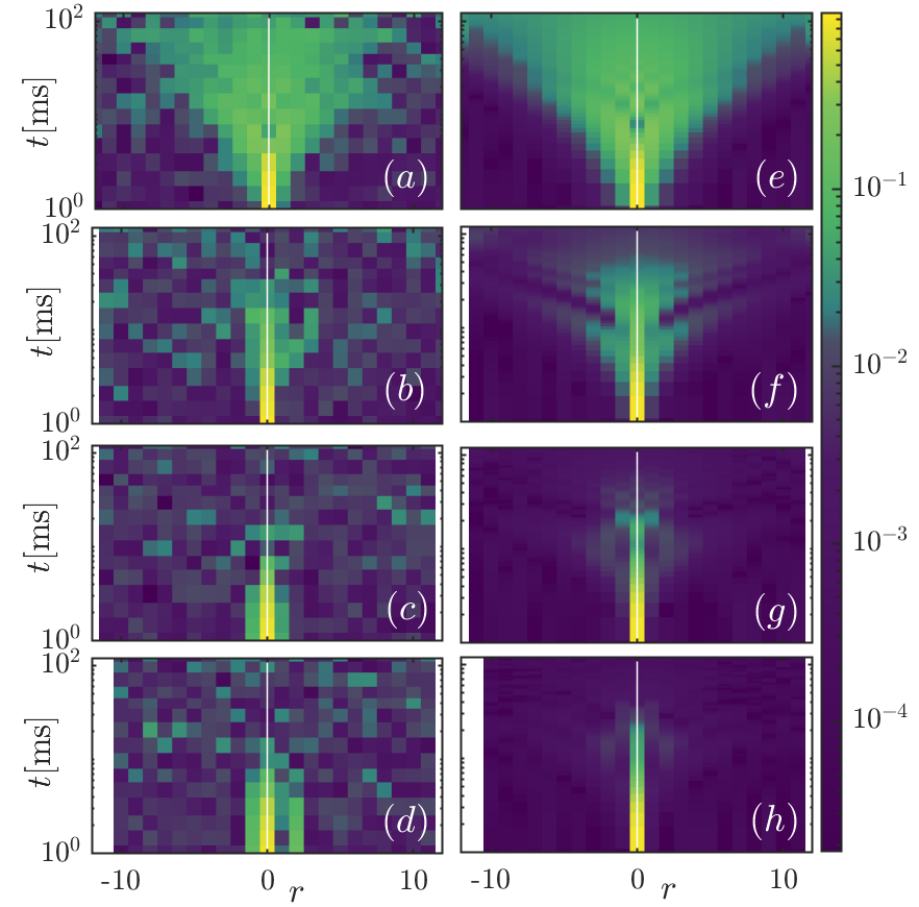}
 \caption{
Spatiotemporal correlation functions $ C_{m}(r,t) = \langle \hat{\mathcal O}_{m,r}(t)\hat{\mathcal{O}}_{m,0} \rangle_c$, for $\mu=m=1,2,3,4$ in panels $(a-d)$, with the corresponding numerical results shown in panels $(e-h)$.
The Hamiltonian parameters are the same as in Fig.~\ref{fig:combined}.  }
\label{fig:profile}
\end{figure}

\paragraph{\textbf{Universal character of the relaxation decay regimes ---}} 
We can also probe the universal character of the relaxation by studying different operators, even with different support, which have the same overlap order. ROI, when saturated, predicts a similar algebraic decay for them. We thus consider the family of operators  
\begin{align}
 \hat{\mathcal O'}_{m,i} = \hat S^z_i \hat S^z_{i+m}, 
\label{Eq:4:Otheroperators}
\end{align} 
where actually $\hat{\mathcal O'}_{1,i}=\hat{\mathcal O}_{2,i}$, each having overlap order $\mu=2$ with $\hat{S}^z$ \footnote{Note that $\hat{\mathcal O'}_{2,i}$ and $\hat{H}$, while $\hat{\mathcal O'}_{3,i}$ have respectively overlap order 4 and 6 with the Hamiltonian $\hat{H}$.}. 
In Fig.~\ref{fig:combined}$(e$-$g)$ we plot the experimental results about the two-time correlators for the corresponding operators $\hat{\mathcal O'}_{m,i}$ and we compare them with the dashed line, proportional to $t^{-1}$, showing a good match. Similarly to Fig.~\ref{fig:combined}$(e)$, we also consider the ratio $\varrho(t)$ in Eq.~(\ref{eq:ratio}), and we observe in Fig.~\ref{fig:combined}$(h)$ a good collapse of the curves. This demonstrates that indeed the autocorrelations of these operators with different support show the same algebraic relaxation.     

\paragraph{\textbf{Spatio-temporal relaxation of the correlations ---}} 
Until now, we have probed the hierarchy of relaxation decays only as a function of time, for operators localized around the same lattice site. However, 
since this dynamics is related to information scrambling and the transport of local conserved quantities, 
we now consider the spatio-temporal correlators $ C_{m}(r,t) = \langle \hat{\mathcal O}_{m,r}(t)\hat{\mathcal{O}}_{m,0} \rangle_c$, where position $0$ is the center site of the system. 
This allows us to probe also how two-time and two-location correlators relax for the same system but for different operators with a different overlap order with the local conserved quantities. 
The behavior of these correlators is depicted by heat maps in Fig.~\ref{fig:profile} both for experimental (left column) and simulated (right column) data. In particular, we have used operators with overlap order varying from $\mu=1$ to $4$, respectively from top to bottom panels. 
The noise in the experimental figures is due both to shot noise and also to sampling noise, as we use only 256 initial conditions, and not $2^L$. 
This figure 
shows that the correlators scramble more rapidly for $\mu=3,\;4$, while for $\mu=2$ it is already possible to observe the survival and spreading of the spatio-temporal correlations, and for $\mu=1$ the correlations show a long-lasting signal over a wide domain. 
This shows clearly that correlations of operators with larger overlap exponent $\mu$ decay much faster than those with smaller one.  


\paragraph{\textbf{Conclusions} ---}

In this article, we have studied the relaxation dynamics of many-body quantum systems with a trapped-ion quantum simulator, with a particular focus on how different observables relax towards equilibrium. 
Building on ETH and hydrodynamics, and for systems with a single local conserved quantity, it was theorized that a universal hierarchy of relaxation regimes would emerge \cite{cwxmp-24}. This is described by the relaxation-overlap inequality (ROI), and it had yet to be experimentally verified.   
Importantly, though, because the system modeled by our quantum simulator has two local conserved quantities, energy and total magnetization, we had to first extend the theory of the relaxation-overlap inequality to include also more conserved quantities. Then, thanks to the precision measurement allowed by the platform, we have experimentally established the existence of a universal hierarchy of algebraic relaxation regimes as expected by the saturated inequality we predict, and we could 
probe 
the more rapid scrambling of the correlations for operators with larger overlap order $\mu$. 
This goes much beyond the experimental verification of the emergence of hydrodynamics \cite{Joshi-22}, where only a class of single-site observables were considered, resulting solely in a single type of relaxation. Interestingly, for our experimental verification the emergence of universal hierarchical relaxation discussed here, we have used data already collected in that experiment, which brings up the interesting point that the same data contains a lot more physical insight.  

Future work could investigate in detail the role of dimensionality of the system, finite temperature and finite magnetization regimes, and the universality of spatio-temporal correlations for $\mu>1$. Furthermore, the theory could be extended to systems with more complex symmetries, such as $SU(2)$ or other non-Abelian ones, and their corresponding conserved quantities. 



\paragraph{\textbf{Acknowledgements} ---}
We acknowledge fruitful discussion with X. Xu. 
This work is supported by the ANR project LOQUST ANR-23-CE47-0006-02 (L.M.~and L.C.) and by the PEPR Dyn-1D ANR-23-
PETQ-0001 (L.M.).
This work is part of HQI (www.hqi.fr) initiative and is supported by France 2030 under the French National Research Agency grant number ANR-22-PNCQ-0002 (L.M.). 
This work is also supported by the Singapore Ministry of Education grant MOE-T2EP50123-0017, and from the Centre for Quantum Technologies grant CQT$\_$SUTD$\_$2025$\_$01 (D.P.). This work is also supported by MOE tier 1 grant number ``SAP 2026$\_$001'' (M.K.J.).

\bibliography{bibliography}

@article{Mori_2018,
   title={Thermalization and prethermalization in isolated quantum systems: a theoretical overview},
   volume={51},
   ISSN={1361-6455},
   url={http://dx.doi.org/10.1088/1361-6455/aabcdf},
   DOI={10.1088/1361-6455/aabcdf},
   number={11},
   journal={Journal of Physics B: Atomic, Molecular and Optical Physics},
   publisher={IOP Publishing},
   author={Mori, Takashi and Ikeda, Tatsuhiko N and Kaminishi, Eriko and Ueda, Masahito},
   year={2018},
   month=may, pages={112001} }

@article{Silva_review_RMP,
  title = {Colloquium: Nonequilibrium dynamics of closed interacting quantum systems},
  author = {Polkovnikov, Anatoli and Sengupta, Krishnendu and Silva, Alessandro and Vengalattore, Mukund},
  journal = {Rev. Mod. Phys.},
  volume = {83},
  issue = {3},
  pages = {863--883},
  numpages = {0},
  year = {2011},
  month = {Aug},
  publisher = {American Physical Society},
  doi = {10.1103/RevModPhys.83.863},
  url = {https://link.aps.org/doi/10.1103/RevModPhys.83.863}
}

@article{Srednicki-99,
  title={The approach to thermal equilibrium in quantized chaotic systems},
  author={Srednicki, Mark},
  journal={Journal of Physics A: Mathematical and General},
  volume={32},
  number={7},
  pages={1163},
  year={1999},
  publisher={IOP Publishing},
  url={https://iopscience.iop.org/article/10.1088/0305-4470/32/7/007}
}

@article{Deutsch-91,
  title={Quantum statistical mechanics in a closed system},
  author={Deutsch, Josh M},
  journal={Phys. Rev. A},
  volume={43},
  number={4},
  pages={2046},
  year={1991},
  publisher={APS},
  doi = {10.1103/PhysRevA.43.2046}
}

@article{berry1977,
  title={Regular and irregular semiclassical wavefunctions},
  author={Berry, Michael V},
  journal={Journal of Physics A: Mathematical and General},
  volume={10},
  number={12},
  pages={2083},
  year={1977},
  publisher={IOP Publishing},
  doi = {10.1088/0305-4470/10/12/016}
}

@article{rdo-08,
  title={Thermalization and its mechanism for generic isolated quantum systems},
  author={Rigol, Marcos and Dunjko, Vanja and Olshanii, Maxim},
  journal={Nature},
  volume={452},
  number={7189},
  pages={854--858},
  year={2008},
  publisher={Nature Publishing Group UK London},
  doi = { https://doi.org/10.1038/nature06838}
}

@article{Rigol-09,
  title = {Breakdown of Thermalization in Finite One-Dimensional Systems},
  author = {Rigol, Marcos},
  journal = {Phys. Rev. Lett.},
  volume = {103},
  issue = {10},
  pages = {100403},
  numpages = {4},
  year = {2009},
  month = {Sep},
  publisher = {American Physical Society},
  doi = {10.1103/PhysRevLett.103.100403},
  url = {https://link.aps.org/doi/10.1103/PhysRevLett.103.100403}
}

@article{sr-10,
  title={Localization and the effects of symmetries in the thermalization properties of one-dimensional quantum systems},
  author={Santos, Lea F and Rigol, Marcos},
  journal={Phys. Rev. E},
  volume={82},
  number={3},
  pages={031130},
  year={2010},
  publisher={APS},
  doi = {10.1103/PhysRevE.82.031130}
}

@article{rs-10,
  title={Quantum chaos and thermalization in gapped systems},
  author={Rigol, Marcos and Santos, Lea F},
  journal={Phys. Rev. A},
  volume={82},
  number={1},
  pages={011604},
  year={2010},
  publisher={APS},
  doi = {10.1103/PhysRevA.82.011604}
}

@article{bkl-10,
  title = {Effect of Rare Fluctuations on the Thermalization of Isolated Quantum Systems},
  author = {Biroli, Giulio and Kollath, Corinna and L\"auchli, Andreas M.},
  journal = {Phys. Rev. Lett.},
  volume = {105},
  issue = {25},
  pages = {250401},
  numpages = {4},
  year = {2010},
  month = {Dec},
  publisher = {American Physical Society},
  doi = {10.1103/PhysRevLett.105.250401},
  url = {https://link.aps.org/doi/10.1103/PhysRevLett.105.250401}
}

@article{krrg-12,
  title = {Quantum quenches in disordered systems: Approach to thermal equilibrium without a typical relaxation time},
  author = {Khatami, Ehsan and Rigol, Marcos and Rela\~no, Armando and Garc\'{\i}a-Garc\'{\i}a, Antonio M.},
  journal = {Phys. Rev. E},
  volume = {85},
  issue = {5},
  pages = {050102},
  numpages = {5},
  year = {2012},
  month = {May},
  publisher = {American Physical Society},
  doi = {10.1103/PhysRevE.85.050102},
  url = {https://link.aps.org/doi/10.1103/PhysRevE.85.050102}
}

@article{ghl-12,
  title = {{Thermalization of local observables in small Hubbard lattices}},
  author = {Genway, S. and Ho, A. F. and Lee, D. K. K.},
  journal = {Phys. Rev. A},
  volume = {86},
  issue = {2},
  pages = {023609},
  numpages = {19},
  year = {2012},
  month = {Aug},
  publisher = {American Physical Society},
  doi = {10.1103/PhysRevA.86.023609},
  url = {https://link.aps.org/doi/10.1103/PhysRevA.86.023609}
}

@article{bmh-14,
  title={Finite-size scaling of eigenstate thermalization},
  author={Beugeling, Wouter and Moessner, Roderich and Haque, Masudul},
  journal={Phys. Rev. E},
  volume={89},
  number={4},
  pages={042112},
  year={2014},
  publisher={APS},
  doi = {10.1103/PhysRevE.89.042112}
}

@article{bmh-15,
  title={Off-diagonal matrix elements of local operators in many-body quantum systems},
  author={Beugeling, Wouter and Moessner, Roderich and Haque, Masudul},
  journal={Phys. Rev. E},
  volume={91},
  number={1},
  pages={012144},
  year={2015},
  publisher={APS},
  doi = {10.1103/PhysRevE.91.012144}
}

@article{kpsr-13,
  title = {Fluctuation-Dissipation Theorem in an Isolated System of Quantum Dipolar Bosons after a Quench},
  author = {Khatami, Ehsan and Pupillo, Guido and Srednicki, Mark and Rigol, Marcos},
  journal = {Phys. Rev. Lett.},
  volume = {111},
  issue = {5},
  pages = {050403},
  numpages = {5},
  year = {2013},
  month = {Jul},
  publisher = {American Physical Society},
  doi = {10.1103/PhysRevLett.111.050403},
  url = {https://link.aps.org/doi/10.1103/PhysRevLett.111.050403}
}

@article{shp-13,
  title = {Eigenstate thermalization within isolated spin-chain systems},
  author = {Steinigeweg, R. and Herbrych, J. and Prelov\ifmmode \check{s}\else \v{s}\fi{}ek, P.},
  journal = {Phys. Rev. E},
  volume = {87},
  issue = {1},
  pages = {012118},
  numpages = {5},
  year = {2013},
  month = {Jan},
  publisher = {American Physical Society},
  doi = {10.1103/PhysRevE.87.012118},
  url = {https://link.aps.org/doi/10.1103/PhysRevE.87.012118}
}

@article{kih-14,
  title = {Testing whether all eigenstates obey the eigenstate thermalization hypothesis},
  author = {Kim, Hyungwon and Ikeda, Tatsuhiko N. and Huse, David A.},
  journal = {Phys. Rev. E},
  volume = {90},
  issue = {5},
  pages = {052105},
  numpages = {8},
  year = {2014},
  month = {Nov},
  publisher = {American Physical Society},
  doi = {10.1103/PhysRevE.90.052105},
  url = {https://link.aps.org/doi/10.1103/PhysRevE.90.052105}
}

@article{skngg-14,
  title = {Pushing the Limits of the Eigenstate Thermalization Hypothesis towards Mesoscopic Quantum Systems},
  author = {Steinigeweg, R. and Khodja, A. and Niemeyer, H. and Gogolin, C. and Gemmer, J.},
  journal = {Phys. Rev. Lett.},
  volume = {112},
  issue = {13},
  pages = {130403},
  numpages = {5},
  year = {2014},
  month = {Apr},
  publisher = {American Physical Society},
  doi = {10.1103/PhysRevLett.112.130403},
  url = {https://link.aps.org/doi/10.1103/PhysRevLett.112.130403}
}

@article{ksg-15,
  title = {Relevance of the eigenstate thermalization hypothesis for thermal relaxation},
  author = {Khodja, Abdellah and Steinigeweg, Robin and Gemmer, Jochen},
  journal = {Phys. Rev. E},
  volume = {91},
  issue = {1},
  pages = {012120},
  numpages = {5},
  year = {2015},
  month = {Jan},
  publisher = {American Physical Society},
  doi = {10.1103/PhysRevE.91.012120},
  url = {https://link.aps.org/doi/10.1103/PhysRevE.91.012120}
}

@article{mfsr-16,
  title = {Eigenstate thermalization in the two-dimensional transverse field Ising model},
  author = {Mondaini, Rubem and Fratus, Keith R. and Srednicki, Mark and Rigol, Marcos},
  journal = {Phys. Rev. E},
  volume = {93},
  issue = {3},
  pages = {032104},
  numpages = {9},
  year = {2016},
  month = {Mar},
  publisher = {American Physical Society},
  doi = {10.1103/PhysRevE.93.032104},
  url = {https://link.aps.org/doi/10.1103/PhysRevE.93.032104}
}

@article{km-63,
  title={Hydrodynamic equations and correlation functions},
  author={Kadanoff, Leo P and Martin, Paul C},
  journal={Annals of Physics},
  volume={24},
  pages={419--469},
  year={1963},
  publisher={Elsevier},
  doi = {10.1016/aphy.2000.6023}
}

@book{Spohn-12,
  title={Large scale dynamics of interacting particles},
  author={Spohn, Herbert},
  year={2012},
  publisher={Springer Science \& Business Media}
}

@article{cdy-16,
  title={Emergent hydrodynamics in integrable quantum systems out of equilibrium},
  author={Castro-Alvaredo, Olalla A and Doyon, Benjamin and Yoshimura, Takato},
  journal={Phys. Rev. X},
  volume={6},
  number={4},
  pages={041065},
  year={2016},
  publisher={APS},
  doi = {10.1103/PhysRevX.6.041065}
}

@article{bcdf-16,
  title={{Transport in out-of-equilibrium XXZ chains: Exact profiles of charges and currents}},
  author={Bertini, Bruno and Collura, Mario and De Nardis, Jacopo and Fagotti, Maurizio},
  journal={Phys. Rev. Lett.},
  volume={117},
  number={20},
  pages={207201},
  year={2016},
  publisher={APS},
  doi = {10.1103/PhysRevLett.117.207201}
}

@article{Doyon-22,
  title={Diffusion and superdiffusion from hydrodynamic projections},
  author={Doyon, Benjamin},
  journal={Journal of Statistical Physics},
  volume={186},
  number={2},
  pages={25},
  year={2022},
  publisher={Springer},
  doi = {10.1007/s10955-021-02863-6}
}

@article{dbd-19,
  title={Diffusion in generalized hydrodynamics and quasiparticle scattering},
  author={De Nardis, Jacopo and Bernard, Denis and Doyon, Benjamin},
  journal={SciPost Physics},
  volume={6},
  number={4},
  pages={049},
  year={2019},
  doi = {10.21468/SciPostPhys.6.4.049}
}

@article{lmmr-14,
  title = {Hydrodynamic long-time tails after a quantum quench},
  author = {Lux, Jonathan and M\"uller, Jan and Mitra, Aditi and Rosch, Achim},
  journal = {Phys. Rev. A},
  volume = {89},
  issue = {5},
  pages = {053608},
  numpages = {8},
  year = {2014},
  month = {May},
  publisher = {American Physical Society},
  doi = {10.1103/PhysRevA.89.053608},
  url = {https://link.aps.org/doi/10.1103/PhysRevA.89.053608}
}

@article{bkv-05,
  title = {How generic scale invariance influences quantum and classical phase transitions},
  author = {Belitz, D. and Kirkpatrick, T. R. and Vojta, Thomas},
  journal = {Rev. Mod. Phys.},
  volume = {77},
  issue = {2},
  pages = {579--632},
  numpages = {0},
  year = {2005},
  month = {Jul},
  publisher = {American Physical Society},
  doi = {10.1103/RevModPhys.77.579},
  url = {https://link.aps.org/doi/10.1103/RevModPhys.77.579}
}

@article{baw-06,
  title = {Nonlinear Quantum Shock Waves in Fractional Quantum Hall Edge States},
  author = {Bettelheim, E. and Abanov, Alexander G. and Wiegmann, P.},
  journal = {Phys. Rev. Lett.},
  volume = {97},
  issue = {24},
  pages = {246401},
  numpages = {4},
  year = {2006},
  month = {Dec},
  publisher = {American Physical Society},
  doi = {10.1103/PhysRevLett.97.246401},
  url = {https://link.aps.org/doi/10.1103/PhysRevLett.97.246401}
}

@article{msf-09,
  title = {Graphene: A Nearly Perfect Fluid},
  author = {M\"uller, Markus and Schmalian, J\"org and Fritz, Lars},
  journal = {Phys. Rev. Lett.},
  volume = {103},
  issue = {2},
  pages = {025301},
  numpages = {4},
  year = {2009},
  month = {Jul},
  publisher = {American Physical Society},
  doi = {10.1103/PhysRevLett.103.025301},
  url = {https://link.aps.org/doi/10.1103/PhysRevLett.103.025301}
}

@article{ngtsm-15,
  title={Hydrodynamics in graphene: Linear-response transport},
  author={Narozhny, BN and Gornyi, IV and Titov, M and Sch{\"u}tt, Michael and Mirlin, AD},
  journal={Phys. Rev. B},
  volume={91},
  number={3},
  pages={035414},
  year={2015},
  publisher={APS},
  doi = {10.1103/PhysRevB.91.035414}
}

@article{ttgp-15,
  title={Nonlocal transport and the hydrodynamic shear viscosity in graphene},
  author={Torre, Iacopo and Tomadin, Andrea and Geim, Andre K and Polini, Marco},
  journal={Phys. Rev. B},
  volume={92},
  number={16},
  pages={165433},
  year={2015},
  publisher={APS},
  doi = {10.1103/PhysRevB.92.165433}
}

@article{mknsm-16,
  title={{Evidence for hydrodynamic electron flow in PdCoO2}},
  author={Moll, Philip JW and Kushwaha, Pallavi and Nandi, Nabhanila and Schmidt, Burkhard and Mackenzie, Andrew P},
  journal={Science},
  volume={351},
  number={6277},
  pages={1061--1064},
  year={2016},
  publisher={American Association for the Advancement of Science},
  doi = {https://doi.org/10.1126/science.aac8385}
}

@article{lf-18,
  title={Hydrodynamics of electrons in graphene},
  author={Lucas, Andrew and Fong, Kin Chung},
  journal={Journal of Physics: Condensed Matter},
  volume={30},
  number={5},
  pages={053001},
  year={2018},
  publisher={IOP Publishing},
  doi = {10.1088/1361-648X/aaa274}
}

@article{Crossno-16,
  title={{Observation of the Dirac fluid and the breakdown of the Wiedemann-Franz law in graphene}},
  author={Crossno, Jesse and Shi, Jing K and Wang, Ke and Liu, Xiaomeng and Harzheim, Achim and Lucas, Andrew and Sachdev, Subir and Kim, Philip and Taniguchi, Takashi and Watanabe, Kenji and others},
  journal={Science},
  volume={351},
  number={6277},
  pages={1058--1061},
  year={2016},
  publisher={American Association for the Advancement of Science},
  doi = {10.1126/science.aad0343}
}

@article{cwxmp-24,
  title = {Hydrodynamics and the Eigenstate Thermalization Hypothesis},
  author = {Capizzi, Luca and Wang, Jiaozi and Xu, Xiansong and Mazza, Leonardo and Poletti, Dario},
  journal = {Phys. Rev. X},
  volume = {15},
  issue = {1},
  pages = {011059},
  numpages = {21},
  year = {2025},
  month = {Mar},
  publisher = {American Physical Society},
  doi = {10.1103/PhysRevX.15.011059},
  url = {https://link.aps.org/doi/10.1103/PhysRevX.15.011059}
}

@article{WangMazza2026,
  title = {Ergodicity and hydrodynamics: From quantum to classical spin systems},
  author = {Wang, Jiaozi and Capizzi, Luca and Poletti, Dario and Mazza, Leonardo},
  journal = {Phys. Rev. E},
  volume = {113},
  issue = {6},
  pages = {064102},
  numpages = {13},
  year = {2026},
  month = {Jun},
  publisher = {American Physical Society},
  doi = {10.1103/mm75-dmt4},
  url = {https://link.aps.org/doi/10.1103/mm75-dmt4}
}

@article{NishikawaSaito2025,
  title = {Energy Diffusion in the Long-Range Interacting Spin Systems},
  author = {Nishikawa, Hideaki and Saito, Keiji},
  journal = {Phys. Rev. Lett.},
  volume = {135},
  issue = {14},
  pages = {147102},
  numpages = {9},
  year = {2025},
  month = {Oct},
  publisher = {American Physical Society},
  doi = {10.1103/hsbt-c46n},
  url = {https://link.aps.org/doi/10.1103/hsbt-c46n}
}

@article{Joshi-22,
   title={Observing emergent hydrodynamics in a long-range quantum magnet},
   volume={376},
   ISSN={1095-9203},
   url={http://dx.doi.org/10.1126/science.abk2400},
   DOI={10.1126/science.abk2400},
   number={6594},
   journal={Science},
   publisher={American Association for the Advancement of Science (AAAS)},
   author={Joshi, M. K. and Kranzl, F. and Schuckert, A. and Lovas, I. and Maier, C. and Blatt, R. and Knap, M. and Roos, C. F.},
   year={2022},
   month=may, pages={720–724}
}

@Article{Belin-22,
	title={{Charged eigenstate thermalization, Euclidean wormholes and global symmetries in quantum gravity}},
	author={Alexandre Belin and Jan de Boer and Pranjal Nayak and Julian Sonner},
	journal={SciPost Phys.},
	volume={12},
	pages={059},
	year={2022},
	publisher={SciPost},
	doi={10.21468/SciPostPhys.12.2.059},
	url={https://scipost.org/10.21468/SciPostPhys.12.2.059},
}

@article{mbisy-23,
  title = {Non-Abelian Eigenstate Thermalization Hypothesis},
  author = {Murthy, Chaitanya and Babakhani, Arman and Iniguez, Fernando and Srednicki, Mark and Yunger Halpern, Nicole},
  journal = {Phys. Rev. Lett.},
  volume = {130},
  issue = {14},
  pages = {140402},
  numpages = {8},
  year = {2023},
  month = {Apr},
  publisher = {American Physical Society},
  doi = {10.1103/PhysRevLett.130.140402},
  url = {https://link.aps.org/doi/10.1103/PhysRevLett.130.140402}
}

@article{PatilRigol_2025,
  title = {Eigenstate thermalization in spin-$\frac{1}{2}$ systems with SU(2) symmetry},
  author = {Patil, Rohit and Rigol, Marcos},
  journal = {Phys. Rev. B},
  volume = {111},
  issue = {20},
  pages = {205126},
  numpages = {14},
  year = {2025},
  month = {May},
  publisher = {American Physical Society},
  doi = {10.1103/PhysRevB.111.205126},
  url = {https://link.aps.org/doi/10.1103/PhysRevB.111.205126}
}

@article{BalachandranPoletti2021,
  title = {From the eigenstate thermalization hypothesis to algebraic relaxation of OTOCs in systems with conserved quantities},
  author = {Balachandran, Vinitha and Benenti, Giuliano and Casati, Giulio and Poletti, Dario},
  journal = {Phys. Rev. B},
  volume = {104},
  issue = {10},
  pages = {104306},
  numpages = {12},
  year = {2021},
  month = {Sep},
  publisher = {American Physical Society},
  doi = {10.1103/PhysRevB.104.104306},
  url = {https://link.aps.org/doi/10.1103/PhysRevB.104.104306}
}

@article{BalachandranPoletti2023,
  title = {Slow relaxation of out-of-time-ordered correlators in interacting integrable and nonintegrable spin-$\frac{1}{2}$ XYZ chains},
  author = {Balachandran, Vinitha and Santos, Lea F. and Rigol, Marcos and Poletti, Dario},
  journal = {Phys. Rev. B},
  volume = {107},
  issue = {23},
  pages = {235421},
  numpages = {10},
  year = {2023},
  month = {Jun},
  publisher = {American Physical Society},
  doi = {10.1103/PhysRevB.107.235421},
  url = {https://link.aps.org/doi/10.1103/PhysRevB.107.235421}
}

@Article{BalachandranPoletti2023b,
AUTHOR = {Balachandran, Vinitha and Poletti, Dario},
TITLE = {Relaxation Exponents of OTOCs and Overlap with Local Hamiltonians},
JOURNAL = {Entropy},
VOLUME = {25},
YEAR = {2023},
NUMBER = {1},
ARTICLE-NUMBER = {59},
URL = {https://www.mdpi.com/1099-4300/25/1/59},
PubMedID = {36673199},
ISSN = {1099-4300},
DOI = {10.3390/e25010059}
}

@article{KranzlYungerHalpern2023,
  title = {Experimental Observation of Thermalization with Noncommuting Charges},
  author = {Kranzl, Florian and Lasek, Aleksander and Joshi, Manoj K. and Kalev, Amir and Blatt, Rainer and Roos, Christian F. and Yunger Halpern, Nicole},
  journal = {PRX Quantum},
  volume = {4},
  issue = {2},
  pages = {020318},
  numpages = {19},
  year = {2023},
  month = {Apr},
  publisher = {American Physical Society},
  doi = {10.1103/PRXQuantum.4.020318},
  url = {https://link.aps.org/doi/10.1103/PRXQuantum.4.020318}
}

@article{LasekYungerHalpern2026,
  title = {Numerical evidence for the non-Abelian eigenstate thermalization hypothesis},
  author = {Lasek, Aleksander and Noh, Jae Dong and LeSchack, Jade and Yunger Halpern, Nicole},
  journal = {Phys. Rev. E},
  volume = {113},
  issue = {5},
  pages = {054140},
  numpages = {19},
  year = {2026},
  month = {May},
  publisher = {American Physical Society},
  doi = {10.1103/j8xb-xfln},
  url = {https://link.aps.org/doi/10.1103/j8xb-xfln}
}

@article{MajidiYungerHalpern2026,
  title = {oncommuting conserved charges in quantum thermodynamics and beyond},
  author = {Majidi, Shayan and Braasch, William F. Jr and Lasek, Aleksander and Upadhyaya, Twesh and Kalev, Amir and Yunger Halpern, Nicole},
  journal = {Nature Review Physics},
  volume = {5},
  pages = {689},
  year = {2023},
  publisher = {Nature Publishing Group},
  doi = {10.1038/s42254-023-00641-9},
  url = {https://doi.org/10.1038/s42254-023-00641-9}
}

@article{DQT-Gemmer,
  title = {Dynamical Typicality of Quantum Expectation Values},
  author = {Bartsch, Christian and Gemmer, Jochen},
  journal = {Phys. Rev. Lett.},
  volume = {102},
  issue = {11},
  pages = {110403},
  numpages = {4},
  year = {2009},
  month = {Mar},
  publisher = {American Physical Society},
  doi = {10.1103/PhysRevLett.102.110403},
  url = {https://link.aps.org/doi/10.1103/PhysRevLett.102.110403}
}

@article{Chebyshev,
  title={Computational methods for simulating quantum computers},
  author={De Raedt, Hans and Michielsen, Kristel},
  journal={arXiv preprint quant-ph/0406210},
  year={2004}
}

@article{WangPappalardi2026,
  title = {Eigenstate Thermalization Hypothesis Correlations via Nonlinear Hydrodynamics},
  author = {Wang, Jiaozi and Mishra, Ruchira and Yang, Tian-Hua and Delacr\'etaz, Luca V. and Pappalardi, Silvia},
  journal = {Phys. Rev. Lett.},
  volume = {136},
  issue = {13},
  pages = {130402},
  numpages = {10},
  year = {2026},
  month = {Mar},
  publisher = {American Physical Society},
  doi = {10.1103/prv4-948b},
  url = {https://link.aps.org/doi/10.1103/prv4-948b}
}

@article{ZuYao2021,
  author  = {Zu, C. and Machado, F. and Ye, B. and Choi, S. and Kobrin, B. and Mittiga, T. and Hsieh, S. and Bhattacharyya, P. and Markham, M. and Twitchen, D. and Jarmola, A. and Budker, D. and Laumann, C. R. and Moore, J. E. and Yao, N. Y.},
  title   = {Emergent hydrodynamics in a strongly interacting dipolar spin ensemble},
  journal = {Nature},
  volume  = {597},
  pages   = {45--48},
  year    = {2021},
  doi     = {10.1038/s41586-021-03763-1}
}

@article{PengCappellaro2023,
  title = {Exploiting disorder to probe spin and energy hydrodynamics},
  author = {Peng, P. and Ye, B. and Yao, N. and Cappellaro, P. },
  journal = {Nature Physics},
  volume = {19},
  pages = {1027},
  year = {2023},
  publisher = {Nature Publishing Group},
  doi = {10.1038/s41567-023-02024-4},
  url = {https://doi.org/10.1038/s41567-023-02024-4}
}

@article{WeiZeher2022,
author = {David Wei  and Antonio Rubio-Abadal  and Bingtian Ye  and Francisco Machado  and Jack Kemp  and Kritsana Srakaew  and Simon Hollerith  and Jun Rui  and Sarang Gopalakrishnan  and Norman Y. Yao  and Immanuel Bloch  and Johannes Zeiher },
title = {Quantum gas microscopy of Kardar-Parisi-Zhang superdiffusion},
journal = {Science},
volume = {376},
number = {6594},
pages = {716-720},
year = {2022},
doi = {10.1126/science.abk2397},
URL = {https://www.science.org/doi/abs/10.1126/science.abk2397},
eprint = {https://www.science.org/doi/pdf/10.1126/science.abk2397}}

@article{ShiFan2024,
  title = {Probing spin hydrodynamics on a superconducting quantum simulator},
  author = {Shi, YH. and Sun, ZH. and Wang, YY. and et al.},
  journal = {Nature Communications},
  volume = {15},
  pages = {7573},
  year = {2024},
  publisher = {Nature Publishing Group},
  doi = {10.1038/s41467-024-52082-2},
  url = {https://doi.org/10.1038/s41467-024-52082-2}
}

@article{WienandBloch2024,
  title = {Emergence of fluctuating hydrodynamics in chaotic quantum systems},
  author = {Wienand, J.F. and Karch, S. and Impertro, A. and Schweize, C. and McCulloch, E. and Vasseur, R. and Gopalakrishnan, S. and Aidelsburger, M. and Bloch, I.},
  journal = {Nature Physics},
  volume = {20},
  pages = {1732},
  year = {2024},
  publisher = {Nature Publishing Group},
  doi = {10.1038/s41567-024-02611-z},
  url = {https://doi.org/10.1038/s41567-024-02611-z}
}

@article{RosenbergRoushan2024,
  author  = {Rosenberg, E. and Andersen, T. I. and Samajdar, R. and Petukhov, A. and Hoke, J. C. and Abanin, D. and others},
  title   = {Dynamics of magnetization at infinite temperature in a Heisenberg spin chain},
  journal = {Science},
  volume  = {384},
  number  = {6691},
  pages   = {48--53},
  year    = {2024},
  doi     = {10.1126/science.adi7877}
}

@article{ZhangPoletti2024,
  author  = {Zhang, Pengfei and Gao, Yu and Xu, Xiansong and Wang, Ning and Dong, Hang and Guo, Chu and others},
  title   = {Emergence of Steady Quantum Transport in a Superconducting Processor},
  journal = {Nature Communications},
  volume  = {15},
  pages   = {10115},
  year    = {2024},
  doi     = {10.1038/s41467-024-54332-9}
}

@misc{fn_overlaplongtime, 
howpublished = {We note here that, while Eq.~(\ref{eq:rel_overlap}) indicates a bound, for chaotic enough systems, the long-time relaxation dynamics will be governed by the slowest-decaying hydrodynamic channel for which there is a nonvanishing overlap}
}

\break 

\begin{center}
\begin{large}
\textbf{End Matter}
\end{large}
\end{center}

\paragraph{\textbf{Computation of the overlaps} ---}
We now discuss the overlap order of the observables considered in the main text. 
The Hamiltonian in Eq.~\eqref{eq:Ising_H} has two extensive conserved quantities, namely the energy 
$\hat H$ and the total magnetization $\hat S^z=\sum_i \hat S_i^z$. Therefore, in the
definition of the overlap order, Eq.~\eqref{eq:overlaps}, one has to consider
connected cumulants involving arbitrary products of the two conserved quantities.
For a local observable $\hat{\mathcal O}_{m,i}$, the relevant overlaps are therefore
\begin{equation}
    \left\langle
    \hat{\mathcal O}_{m,i}\,
    \left(\hat H\right)^p
    \left(\hat S^z\right)^q
    \right\rangle_c,
    \qquad
    p,q\geq 0 .
    \label{eq:EM_mixed_overlap}
\end{equation}
The overlap order is determined by the smallest total order $n=p+q$ for which at
least one of the cumulants in Eq.~\eqref{eq:EM_mixed_overlap} is nonzero.

In the experiment we consider the family of observables
\begin{equation}
    \hat{\mathcal O}_{m,i}
    =
    \hat S_i^z \hat S_{i+1}^z\cdots \hat S_{i+m-1}^z,
    \qquad
    m=1,\ldots,4 .
    \label{eq:EM_observables}
\end{equation}
At infinite temperature and zero total magnetization, all mixed overlaps of total
order smaller than $m$ vanish,
\begin{equation}
    \left\langle
    \hat{\mathcal O}_{m,i}\,
    \left(\hat H\right)^p
    \left(\hat S^z\right)^q
    \right\rangle_c
    =
    0,
    \qquad
    p+q<m .
    \label{eq:EM_lower_mixed_overlap}
\end{equation}
At total order $m$, however, there are nonzero overlaps. The pure magnetization
overlap is always nonzero,
\begin{equation}
    \left\langle
    \hat{\mathcal O}_{m,i}
    \left(\hat S^z\right)^m
    \right\rangle_c
    \neq 0 .
    \label{eq:EM_Sz_leading_overlap}
\end{equation}
Thus the overlap order of $\hat{\mathcal O}_{m,i}$ is
\begin{equation}
    \mu(\hat{\mathcal O}_{m,i})=m .
    \label{eq:EM_mu_final}
\end{equation}

We now specify which energy and mixed energy--magnetization overlaps can appear at
this leading order. 
Since $\hat{\mathcal O}_{m,i}$ and
$\hat S^z$ are diagonal in the $\hat S_i^z$ basis, while a single power of $\hat H$
is off-diagonal in this basis, any nonzero infinite-temperature trace must contain
an even number of powers of $\hat H$. Therefore the leading mixed overlaps obey the
selection rule
\begin{equation}
    p+q=m,
    \qquad
    p=0,2,4,\ldots .
    \label{eq:EM_selection_rule}
\end{equation}
Equivalently, at the leading order $m$, the allowed nonzero overlaps are
\begin{equation}
    \left\langle
    \hat{\mathcal O}_{m,i}
    \hat H^p
    \left(\hat S^z\right)^{m-p}
    \right\rangle_c
    \neq 0,
    \qquad
    p=0,2,4,\ldots,m .
    \label{eq:EM_allowed_overlaps}
\end{equation}
Here the last value $p=m$ is present only when $m$ is even. The explicit leading
overlaps for the four observables considered in the main text are summarized in
Table~\ref{tab:EM_overlaps}.

\begin{table}[t]
\centering
\renewcommand{\arraystretch}{1.35}
\begin{tabular}{c c c}
\hline
Observable & $\mu$ & Nonzero leading overlaps \\
\hline
$\hat{\mathcal O}_{1,i}$
&
$1$
&
$\left\langle \hat{\mathcal O}_{1,i} \hat S^z\right\rangle_c$
\\
$\hat{\mathcal O}_{2,i}$
&
$2$
&
$\left\langle \hat{\mathcal O}_{2,i}\left(\hat S^z\right)^2\right\rangle_c$,
$\left\langle \hat{\mathcal O}_{2,i}\left(\hat H\right)^2\right\rangle_c$
\\
$\hat{\mathcal O}_{3,i}$
&
$3$
&
$\left\langle \hat{\mathcal O}_{3,i}\left(\hat S^z\right)^3\right\rangle_c$,
$\left\langle \hat{\mathcal O}_{3,i}\left(\hat H\right)^2\hat S^z\right\rangle_c$
\\
$\hat{\mathcal O}_{4,i}$
&
$4$
&
$\left\langle\! \hat{\mathcal O}_{4,i}\left(\hat S^z\right)^4\right\rangle_{\!\! c}\!$,
$\left\langle\! \hat{\mathcal O}_{4,i}\left(\hat H\right)^2\left(\hat S^z\right)^2\right\rangle_{\!\! c}\!$,
$\left\langle\! \hat{\mathcal O}_{4,i}\left(\hat H\right)^4\right\rangle_{\!\! c}$
\\
\noalign{\vskip 0.1cm}
\hline
\end{tabular}
\caption{
Leading connected overlaps of the observables $\hat{\mathcal O}_{m,i}$ with the
conserved quantities $\hat H$ and $\hat S^z$ at infinite temperature and zero
total magnetization. The overlap order is the smallest total order $p+q$ for which
$\langle \hat{\mathcal O}_{m,i} \hat H^p(\hat S^z)^q\rangle_c$ is nonzero. For the
XX Hamiltonian with $B_i=0$, only even powers of $\hat H$ can contribute to the
leading overlap.
}
\label{tab:EM_overlaps}
\end{table}

The presence of several leading overlaps means that, in principle, several
hydrodynamic channels can contribute to the late-time relaxation. For
$\hat{\mathcal O}_1$, the leading overlap is purely in the magnetization sector.
For $\hat{\mathcal O}_2$, $\hat{\mathcal O}_3$, and $\hat{\mathcal O}_4$, mixed
energy--magnetization overlaps may also appear at the same total overlap order.
Therefore, the asymptotic tail is determined by the slowest hydrodynamic channel
among the conserved modes that enter the leading overlaps. This motivates a direct
comparison between the transport of magnetization and energy.

\begin{figure}[tb]
\includegraphics[width=\columnwidth]{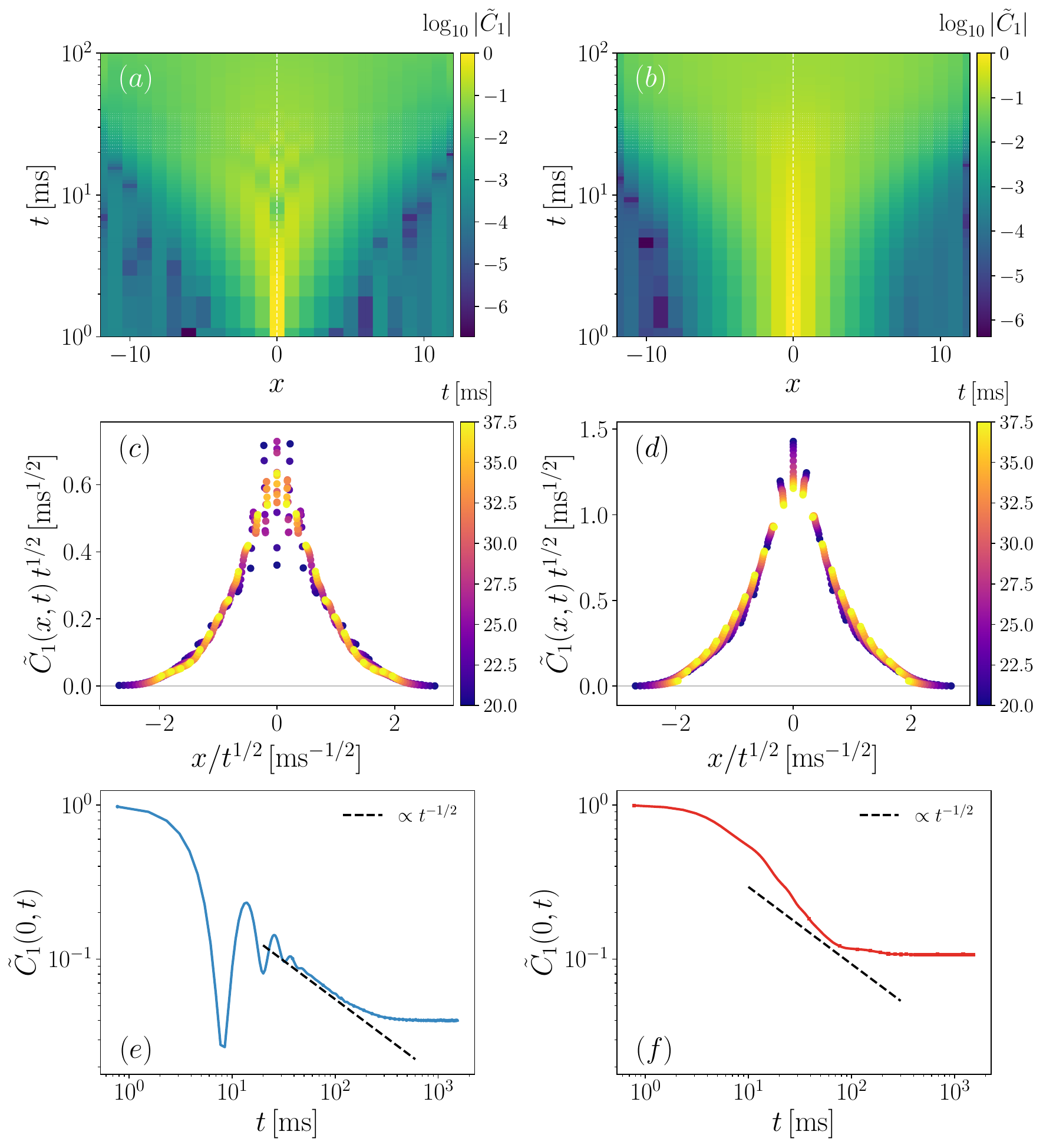}
\caption{
Infinite-temperature correlation functions $ \tilde{C}_{1}(x,t)=\langle\hat{\mathcal{O}}_{1,x}(t)\hat{\mathcal{O}}_{1,0}\rangle_{c}/\langle\hat{\mathcal{O}}_{1,0}^{2}\rangle_{c}$ computed in the full Hilbert space for the local magnetization density $\hat{\mathcal O}_{1,i}=\hat S_i^z$ [panels (a,c,e)] and the local energy density $\hat{\mathcal O}_{1,i}=\hat h_i$, with $\hat h_i$ defined in Eq.~\eqref{eq:EM_local_energy_density}, panels [(b,d,f)]. 
Panels (c,d) display the scaling $ \tilde{C}_{1}(x,t) t^{1/2}$ vs $x/t^{1/2}$. 
Panels (e,f) show $\tilde{C}_{1}(0,t)$ vs $t$, where dashed lines indicate the scaling $\sim t^{-1/2}$.
The Hamiltonian parameters are the same as in Fig.~\ref{fig:combined}.
}
\label{fig:E1}
\end{figure}

For the long-range XX Hamiltonian in Eq.~(\ref{eq:Ising_H}), we define the local energy density as
\begin{equation}
    \hat h_i
    =
    \frac12\sum_{j\neq i}
    J_{ij}
    \left(
    \hat S_i^x\hat S_j^x+\hat S_i^y\hat S_j^y
    \right),
    \label{eq:EM_local_energy_density}
\end{equation}
so that
$
    \hat H=\sum_i \hat h_i .
$
In Fig.~\ref{fig:E1} we study the infinite-temperature correlation functions
$ C_{1}(x,t) = \langle \hat{\mathcal O}_{1,x}(t)\hat{\mathcal{O}}_{1,0} \rangle_c$, which we have numerically evaluated, for the operators 
$    \hat{\mathcal O}_{1,i}=\hat S_i^z$ and $ \hat h_i$. 
In particular, we consider the rescaled quantity
\begin{equation}
    \tilde{C}_{1}(x,t)=\dfrac{\langle\hat{\mathcal{O}}_{1,x}(t)\hat{\mathcal{O}}_{1,0}\rangle_{c}}{\langle\hat{\mathcal{O}}_{1,0}^{2}\rangle_{c}}
\end{equation}
and in panels (c,d) we show the emergence of a transport phenomenon, with the initial perturbation spreading through the whole system.
In order to quantify the nature of this transport, in panels (a,b) we display a rescaling of the correlation function $\tilde C_{1}(x,t) t^{1/2}$ vs $x/t^{1/2}$, which is expected to collapse on a single line for normal diffusive transport. 
The collapse for $\hat{\mathcal O}_{1,i}=\hat S_i^z$, panel (a), and for $\hat{\mathcal O}_{1,i}=\hat h_i$, panel (b), 
is very clear when looking at the tails of the distribution, whereas the data at $x=0$ in both panel are not yet perfectly collapsed, which we impute to finite-time effects. 
Both plots are consistent with normal diffusive transport, in agreement with previous results for the same long-range XX model at $\alpha=3/2$. 
%

Having established the diffusive nature of the transport, we focus in panels (e,f) on the finite-time effects of the equal-space autocorrelation functions, $ \tilde C_{1}(0,t)$, for the magnetization density (blue line in panel (e)) and energy density (red line in panel (f)); the dashed lines, indicating the normal diffusive scaling $t^{-1/2}$ are guides to the eye. 
The energy-density correlation does not display a decay slower than the magnetization autocorrelation. Therefore, for the parameters considered here, the energy channel is not expected to dominate the late-time relaxation. 
The observed hierarchy is thus controlled by the normal diffusive magnetization channel, leading to
\begin{equation}
    C_m(t)\sim t^{-m/2}.
    \label{eq:EM_hierarchy}
\end{equation}

\section{Evaluation of expectation values}
The expectation values of the two-time correlators 
\begin{align}
    C(t)&=\braket{\hat{\mathcal{O}}(t) \hat{\mathcal{O}} (0)}_{T=\infty} 
\end{align}
is given, for on operator given by the tensor product of identity matrices and $\hat{S}^z_i$, i.e. $\hat{\mathcal{O}}(0)=\prod_{i\in \mathcal{D}} \hat{S}^z_i$ over a domain $\mathcal{D}$, by 
\begin{align}
    C(t)&= \frac{1}{Z} \sum_{ s_{-\frac{\Ll}{2}}, \dots, s_{\frac{\Ll}{2}}} \braket{ s_{-\frac{\Ll}{2}}\dots s_{\frac{\Ll}{2}}|\hat{\mathcal{O}}(t) \hat{\mathcal{O}}(0)| s_{-\frac{\Ll}{2}}\dots s_{\frac{\Ll}{2}}} \nonumber \\
    &=\frac{1}{Z} \sum_{ s_{-\frac{\Ll}{2}}, \cdots, s_{\frac{\Ll}{2}}} \left(\prod_{i\in\mathcal{D}} s_i \right) \nonumber \\ 
    &\;\;\;\;\;\;\;\;\;\;\;\;\;\;\times \braket{s_{-\frac{\Ll}{2}}\cdots s_{\frac{\Ll}{2}}| \left(\prod_{i\in\mathcal{D}}\hat{S}^z_i(t) \right)| s_{-\frac{\Ll}{2}}\dots s_{\frac{\Ll}{2}}} \label{eq:infinite_T_allprod} ,
\end{align}
where $\sigma_j = \pm 1/2$, which reduces the evaluation of a two-time correlation function to the weighted sum of single-time functions using product initial states. 
Here, we used $\Ll=L-1$, as for us $L=25$. 
However, rather than consider all possible configurations as there are exponentially large number of samples, we sample $60$ configurations from the sector with total magnetization $\sum_l s_l = +1/2$ (where the center site fixed in the configuration $+1/2$). In the subsequent set of $60$, we choose configurations with polarity opposite for all spins except the middle site. On top of this, we consider another set of configurations ($60+60$), where the total magnetization of the system is $-1/2$. This results in a total of $N_u=240$ initial state configurations.  

\subsection{Standard deviation of mean} The error bars are calculated from the standard error propagation formula used to estimate mean observable expectation values from all configuration settings and the measurement samples. The experimental measurements were taken with a total of $N_m = 50$ samples and $N_u =240$ initial state configurations and $N_t=16$ time points. A slight change in the measurement budget was considered for $t=70, 100$ and $120$ ms time settings due to experimental restrictions, where only $N_m = 25$ samples were recorded. The standard error of the mean for an arbitrary observable expectation value $\langle \mathcal{O} \rangle$ is expressed as
\begin{equation}
    \sigma_{std} = \sqrt{\frac{1}{N_m^2 N_u^2}\sum_{j=1}^{N_u}\sum_{i=1}^{N_m}(x_{i,j}-\mu)^2}
\end{equation}
where $x_{ij} = s^{(i,j)}_ks_{k+1}^{(i,j)}...s_{k+d}^{(i,j)}$ for the $j^{th}$ configuration and the $i^{th}$ experimental sample. In case where all possible combinations arising from different sites ($k$) were averaged, an additional $1/\sqrt{n}$ scaling was used to calculate the standard error of the mean.

\subsection{State preparation and measurement error} In the present study, we evaluated data taken for the manuscript \cite{Joshi-22}, with an emphasis on extended observables estimated from full bit strings. In the previous study, only single-site observables were used in contrast to the present studies. The state-preparation and measurement errors are already discussed in the SM of the original manuscript \cite{Joshi-22}. The full procedure for recording the data can also be found in the manuscript. 
\subsection{Details on numerical simulations}

The numerical results presented in the main text and End Matter are obtained using
dynamical quantum typicality (DQT)~\cite{DQT-Gemmer}.
Let us consider a normalized Haar-random state $|\psi\rangle$ in a Hilbert space
of dimension $D$. According to DQT, the two-time correlation function can be evaluated as 
\begin{equation}
    \frac{1}{D}\mathrm{Tr}\!\left[\hat A(t)\hat B\right]
    =
    \langle\psi|\hat A(t)\hat B|\psi\rangle
    +\epsilon(t),
\end{equation}
where 
\begin{equation}
    \epsilon(t)\sim\frac{1}{\sqrt{D}}.
\end{equation}
The accuracy can be further improved by averaging over $N_p$ independent
realizations of the random state, in which case
\begin{equation}
    \epsilon(t)\sim\frac{1}{\sqrt{D N_p}}.
\end{equation}

Nevertheless, for all numerical results presented in this work, we consider a single random state ($N_p=1$) in each Hilbert space considered: in Fig.~\ref{fig:combined}, the calculations are performed in the two magnetization sectors $S^z_{\mathrm{tot}}=\pm1$; in Figs.~\ref{fig:profile} and~\ref{fig:E1}, they are performed in the full Hilbert space.
The real-time evolution of the random states is calculated using a
Chebyshev-polynomial expansion of the time-evolution operator~\cite{Chebyshev}.

\end{document}